\documentclass[twocolumn,letterpaper]{IEEEAerospaceCLS}  

\usepackage[]{graphicx}    
\usepackage[acronym]{glossaries}
\usepackage{url}

\usepackage{graphicx,color}
\usepackage{mathtools}
\usepackage[caption=false]{subfig}
\usepackage{latexsym,amssymb}
\usepackage{enumitem}
\usepackage{tikz}
\usepackage{xspace}

\usepackage{multirow}

\usepackage{url} 

\newcounter{tempctr}
\newcommand{\breakenumistart}{%
  \setcounter{tempctr}{\value{enumi}}%
  \end{enumerate}%
}
\newcommand{\breakenumiend}{%
  \begin{enumerate}%
  \setcounter{enumi}{\value{tempctr}}%
}

\newcommand{\fig}[2][]{Figure~\ref{fig:#2}{#1}}

\newcommand{\secn}[1]{Section~\ref{secn:#1}}

\newcommand{\mdash}{---}

\newcommand{\cf}[1][\ ]{cf.#1}
\newcommand{\ie}[1][\ ]{i.e.#1}
\newcommand{\etc}[1][\ ]{etc.#1}
\newcommand{\eg}[1][\ ]{e.g.#1}
\newcommand{\wrt}[1][\ ]{w.r.t.#1}

\newcommand{\true}{\ensuremath{\mathit{true}}}

\newcommand{\cA}{{\ensuremath{\mathcal{A}}}}

\newcommand{\cC}{{\ensuremath{\mathcal{C}}}}
\newcommand{\cD}{{\ensuremath{\mathcal{D}}}}

\newlength{\templength}

\usepackage{soul}
\soulregister\cite7
\soulregister\ref7
\soulregister\pageref7
\soulregister{\em}{0}
\soulregister{\mdash}{7}
\soulregister{\st}{7}
\soulregister{\bf}{0}
\soulregister{\fig}{7}
\soulregister{\ie}{0}
\soulregister{\etc}{0}
\soulregister{\eg}{0}
\soulregister{\wrt}{0}
\soulregister{\resp}{0}
\soulregister{\cf}{0}

\usepackage[colorinlistoftodos,bordercolor=white]{todonotes}

\graphicspath{{./}{graphics/}{fig/}{../fig/}}

\newcommand{\ignore}[1]{}  

\newacronym{cots}{COTS}{Commercial Off-The-Shelf} 
\newacronym{BIP}{BIP}{Behaviour-Interaction-Priority}
\newacronym{RISD}{RISD}{Rigorous System Design Laboratory}
\newacronym{CDMS}{CDMS}{Control and Data Management Subsystem}
\newacronym{ADCS}{ADCS}{Attitude Determination and Control Subsystem}
\newacronym{ECSS}{ECSS}{European Cooperation for Space}
\newacronym{TRL}{TRL}{Technology Readiness Level}
\newacronym{SSIP}{SSIP}{Swiss Space Implementation Plan} 
\newacronym{osra}{OSRA}{On-Board Software Reference Architecture}

\begin{document}
\title{Robust Software Development for University-Built Satellites}

\author{%
Anton B. Ivanov\\ 
Space Engineering Center, EPFL\\
PPH 334, Station 13\\
1015 Lausanne, Switzerland\\
+41 21 693 6978\\
anton.ivanov@epfl.ch
\and 
Simon Bliudze\\
EPFL IINFCOM LCA2\\
INJ 340, Station 14\\
1015 Lausanne, Switzerland\\
+41 21 693 1397\\
simon.bliudze@epfl.ch
\thanks{\footnotesize 978-1-5090-1613-6/17/$\$31.00$ \copyright2017 IEEE}              
}

\maketitle

\thispagestyle{plain}
\pagestyle{plain}

\begin{abstract}
Satellites and other complex systems now become more and more software
dependent. Even nanosatellites have complexity that can be compared
to scientific instruments launched to Mars. COTS components and subsystems may now be purchased to support payload development. 
On the contrary, the software has to be adapted to the
new payload and, consequently, hardware architecture selected for the satellite.  There is not a
rigorous and robust way to design software for CubeSats or small satellites yet. In this paper, we will briefly review some existing systems and present our approach, which based on Behaviour-Interaction-Priority (BIP) framework. We will describe our experience in implementing fight software simulation and testing in the Swiss CubETH CubeSat project. We will conclude with lessons learned and future utilization of BIP for hardware testing and simulation. 
\end{abstract}

\setcounter{tocdepth}{1}
\tableofcontents

\section{Introduction}
\label{secn:intro}

Flight software is rewritten for each satellite project, despite the existing heritage, due to changed requirements, new payload or updated hardware components. There is not yet a rigorous and robust way to design software and adapt for changes for small satellites yet or CubeSats in a university setting. There are design practices and considerable experience that exist in all major space organisations such as NASA and ESA, however, they are not available to student teams.  To our knowledge, considerable number flight software programs for university satellites is written in C or C++ code. Recent efforts at Vertmont Tech~\cite{Brandon2013} are using industry standard Ada Code. SysML
\footnote{SysML is an extension of the Unified Modeling Language (UML) using UML's profile mechanism} 
can be used to describe the system as a whole~\cite{sysmlpaper} and then check some properties such as energy consumption. 
SysML can be a valid tool for system engineering as a whole, but it is not rigorous enough to allow automatic software behaviour verification and validation. Another approach~\cite{Dathathri2016} is to utilise TuLiP (Temporal Logic Planning Toolbox) with the JPL SCA (Statechart Autocoder) to enable the automatic synthesis of low-level implementation code directly from formal specifications. 

Here we present our approach using the \gls{BIP} framework~\cite{esst4bip},  a component-based language which can be used to build correct-by-construction applications. It has been developed by the Verimag laboratory in Grenoble university and is currently used in the EPFL by the Rigorous System Design laboratory (RISD). \gls{BIP} can be used to formally model complex systems and provides a toolset for their verification and validation and for code generation. The framework is young; therefore there are not many practical designs with it yet. 

Modular nature of the \gls{BIP} approach allows iterative design for satellites in development and adaptation to hardware changes. In order to accomplish modularity, common patterns in components and structures have to be developed. The verification of the correctness in the \gls{BIP} model is made by using a set of common rules in the assembly of atoms and compounds. 

We have used the \gls{BIP} approach in the our CubeSat project (CubETH) to design logic for the operation of a satellite and compiled into machine code, which was executed on the on-board computer.  Software running Control and Data Management system compiled for Cortex-A3 processor design. This work has proven technical feasibility and exposed a number of problems with the approach. Our next goal is to develop a visual environment to provide graphical user interface to simplify generation of the \gls{BIP} code. 

In this work we will present main principles of the \gls{BIP} framework, our implementation, and challenges for implementation on a CubeSat platform. 

\section{Motivation} 
\label{secn:motivation}
\subsection*{CubETH CubeSat project} 
CubETH is a cooperative Swiss CubeSat mission \cite{Ivanov2015} to demonstrate new technologies in the area of Global Navigation Satellite System (GNSS)-based n
avigation and the usage of COTS components. The satellite will carry five GNSS patch antennas, each connected to two independent u-blox multi-GNSS receivers. These very small, commercially available low-cost receivers are able to track single-frequency code and phase data of all the major GNSS, i.e. GPS, GLONASS, QZSS, Beidou (and ready for Galileo). Four main science objectives have been defined for the CubETH mission: (1) precise orbit determination using low-cost GNSS receivers, (2) attitude determination based on very short baselines, (3) comparison of the performance in space between GPS and GLONASS (and possibly other GNSS) as an important step for further developments of space-borne GNSS receivers, (4) additional experimental measurements (e.g. for air density estimation during re-entry). The satellite will carry 3 retro-reflectors to enable satellite laser ranging for performance validation of the precise orbit determination. Novel CubeSat technologies \cite{Rossi20151513,Rossi20151493} will also be tested. We will take into account lessons learned from the SwissCube mission, which has been in operation since 2009. A new modular design for the structure as well as connectors will be flown. We also intend to demonstrate the use of a miniaturized low-power command and data handling system, which shall control the satellite.

\begin{figure}[htbp] 
	\centering \includegraphics[width=8cm]{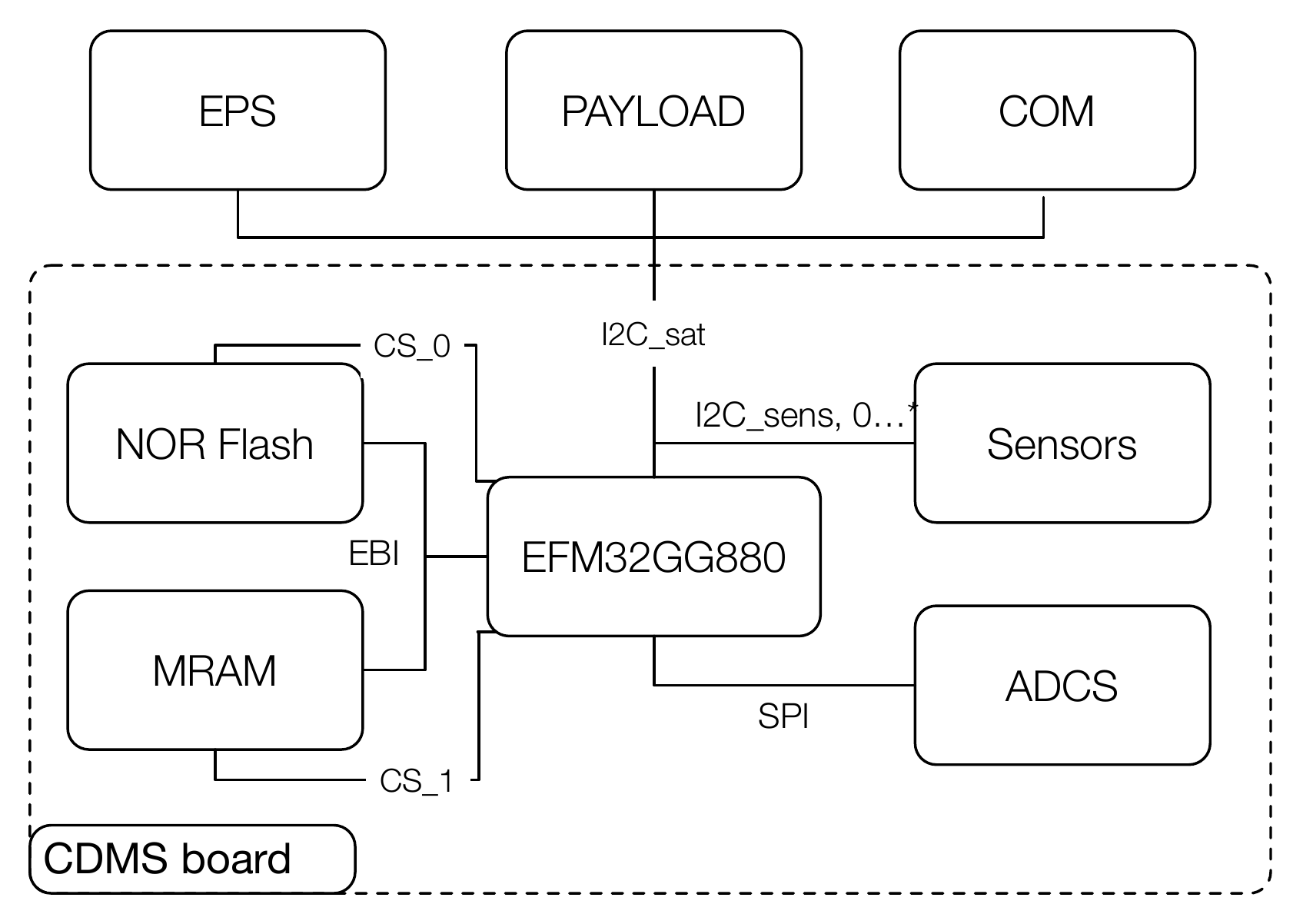}
	\caption{Simplified block-diagram of the CubETH flight software structure, which is centered on the Control and Data Management System}
	\label{CDMSSimplified}
\end{figure}

The components that were most critical for the successful operation of the \gls{CDMS} are the microcontroller and the memories, which are both COTS (Commercial Off-The-Shelf) components. Consequently, components that have already been flown in CubeSats or other spacecraft have been selected.
The \gls{CDMS} will use a microcontroller from the Giant Gecko family by Energy Micro, a Cortex-M3 running at 48 MHz and with a very low power consumption (limited at 36.5 mW).
The inter-board communication protocol is I2C. The \gls{CDMS} (simplified diagram is shown on Figure \ref{CDMSSimplified}) acts as master and requests flight data from the other subsystems. At this stage of the development [15], relevant data to gather is typically housekeeping parameters, as board current consumption, temperature, status flags, or battery voltage. Such data acquisition and I2C communication were implemented on all present subsystems. 

One key property of the \gls{BIP} framework is the capability to automatically generate C++ code, which can be further compiled on a microprocessor platform. For the CubETH project we were able to create a full model of the spacecraft and create toolchain to compile code for running on Cortex-M3 architecture.  

\begin{figure}[htbp] 
	\centering \includegraphics[width=8cm]{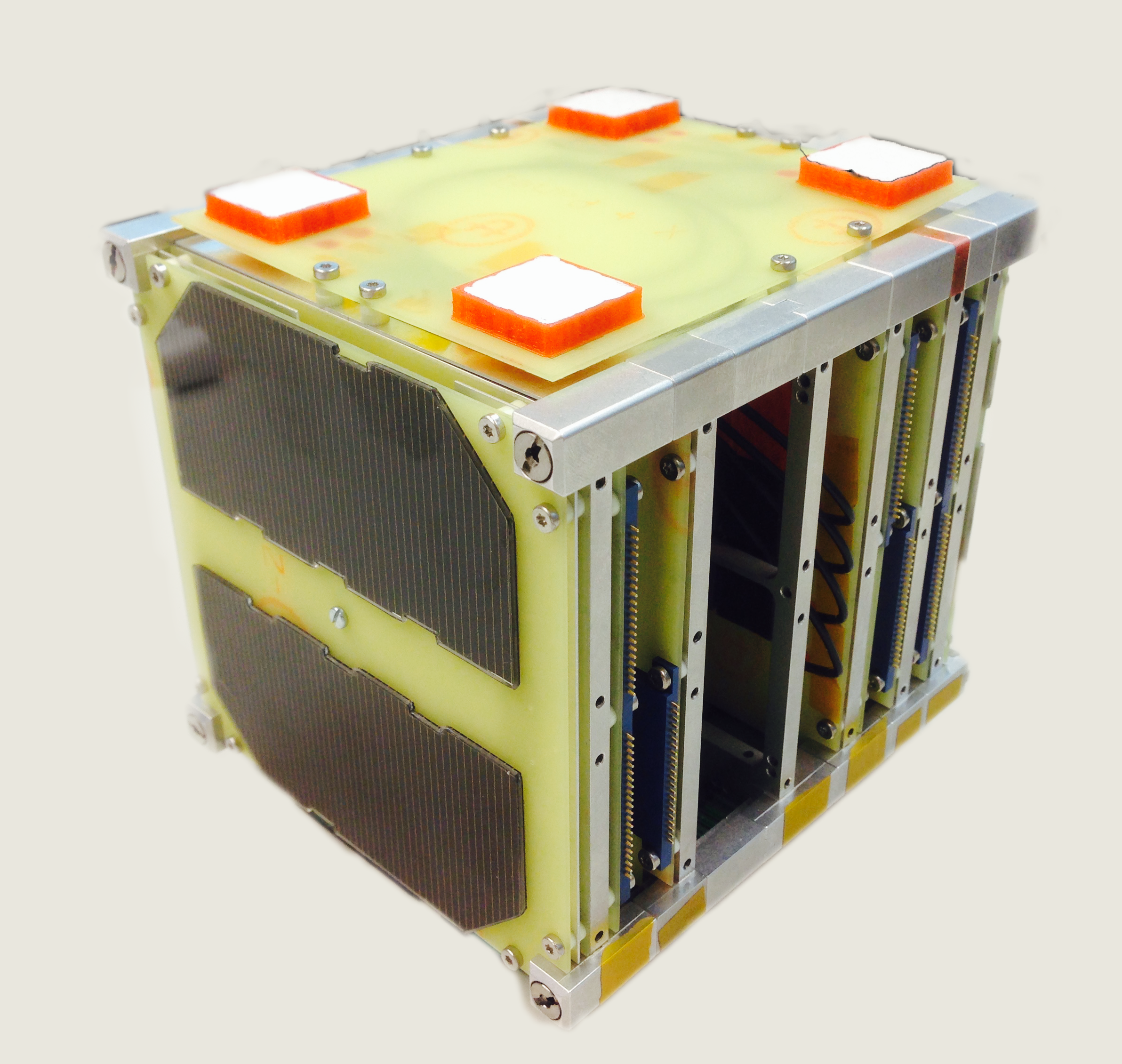}
	\caption{Structural and Thermal model of the CubETH satellite. This is a 1 unit CubeSat measuring 10x10x10 cm.  On the top panel are 4 GNSS antennae for the experiment. Internal electronics are organized in layers.  }
	\label{fig:CubETHPhoto}
\end{figure}

\subsection{Flight software quality attributes}
In the international standards for architecture description and
software engineering a number of quality attributes is specified as
expression of ``ilities'' - qualities such as reliability, portability,
modularity and others. Software designed with these qualities in mind
is more robust to faults in hardware. At the European level, these attributes are
described in a set of \gls{ECSS} standards, such as
\cite{ECSSEST40C}. At NASA, in addition to standard design practices
a recent effort~\cite{Wilmot2016} has been underway to specify a large table of quality
attributes for mission flight software. In this work, we would like to
concentrate primarily on robustness, safety, modularity and
portability qualities for flight software architectures for nano and
mico satellites and scientific instruments. These two broad categories
are the main target customers of the work described in this
proposal.

It is clear that the topic of ``correct-by-construction'' \footnote{"correct-by-construction" implies that all required functionality will be delivered and the correct behavior will exhibited by the compiled code. In other words, many errors in the software can be caught at the compiling time. } software
development is now very actively studied by many groups around the
world to address increased complexity of
modern systems. Our approach is to the use the formal validation framework to create
and verify the system logic,  followed by porting C++ code
directly onto the microprocessor. This approach will ensure the following
``ilities'' of the overall system (following~\cite{Wilmot2016}):

\subsubsection{Reliability} 
The developer will have to focus first on the logic
of the overall system. Before starting on hardware drivers, it will be
possible to model behaviour of the overall system using established
design patterns. The system will also be validated to be compliant
with a number of design rules. Thus, the system will not be 100\%
proof, but it will allow eliminating a number of errors or
ambiguities that are usually recovered late in the project
development cycle. 

\subsubsection{Modularity} 
The system will allow verifying individual
components separately, before integration of the whole system. For
example, temperature sensors can be integrated as ``hardware in the
loop'' to verify logic of operations without connecting them to the
rest of the system. 

\subsubsection{Portability}
The overall system will have a hardware interface
in the form of C or C++ low-level functions, which are implemented to
reflect capabilities of selected components. In the case of component
replacement we need only to replace the corresponding driver, while
keeping overall logic of the system in place.
This would be a major improvement on the current state-of-practice, where  a
change of components also requires modifications of the logic. 
Such changes are extremely costly, once the logic is
implemented already in C++ code and require a lot of reverse
engineering work. 

\section{The BIP framework}
\label{secn:bipframework}

Our approach relies on the \gls{BIP} framework~\cite{main_bip} for
component-based design of correct-by-construction
applications.  \gls{BIP} provides a simple, but powerful mechanism
for the coordination of concurrent components by superposing
three layers: Behaviour, Interaction, and Priority.  First,
component \emph{behaviour} is described by Labelled
Transition Systems (LTS) having transitions labelled with
\emph{ports} and extended with data stored in local
variables.  Ports form the interface of a component and are
used to define its interactions with other components.  They
can also export part of the local variables, allowing access
to the component's data.  Second, \emph{interaction models},
\ie sets of interactions, define the component coordination.
Interactions are sets of ports that define allowed
synchronisations between components.  An interaction model
is defined in a structured manner by using
connectors~\cite{BliSif08-acp-tc}.  Third, \emph{priorities}
are used to impose scheduling constraints and to resolve
conflicts when multiple interactions are enabled
simultaneously.  Interaction and Priority layers are
collectively called \emph{Glue}.

The strict separation between behaviour\mdash \ie stateful
components\mdash and coordination\mdash \ie stateless
connectors and priorities\mdash allows the design of modular
systems that are easy to understand, test and maintain.

The \gls{BIP} language has been implemented as a coordination
language for C++~\cite{main_bip} and Java~\cite{MiSE14p25}.
The core of the framework provides automatic,
semantics-preserving C++ code generation through a tool
structured into:
\begin{description}
	\item[-] a front-end, for parsing a program in the \gls{BIP} {\em
		language} (a BIP-specific extension of C++) and building
	the corresponding \gls{BIP} {\em model} (a language-independent
	model in the Eclipse Modelling Framework);
	
	\item[-] a set of filters for model transformations;
	
	\item[-] and the back-end for the generation of code suitable
	for compiling onto the target platform.
\end{description}
The tool is modular: both the front- and the back-end can be
independently replaced to provide parsing and generating code in other
languages than \gls{BIP} (for input) and C++ (for output).
Execution of the \gls{BIP} model is driven by the \gls{BIP} Engine, which
implements the \gls{BIP} operational semantics.  It is provided as a
precompiled library, linked with the generated C++ code of the model.

The \gls{BIP} framework provides several verification tools: DFinder~\cite{dfinder} for
compositional deadlock-freedom analysis, Kratos (developed in collaboration with FBK\footnote{%
	Fondazione Bruno Kessler: \url{http://www.fbk.eu/}.
})~\cite{esst4bip} for state-reachability checking and BIP-to-NuSMV
tool~\cite{esst4bip}, which transforms \gls{BIP} models into the input
format of the nuXmv model checker, for the
verification of temporal logic properties.

\subsection*{Main features of BIP}
\label{secn:features}

Below we further discuss the main features of the BIP
framework.\footnote{
	A complete \gls{BIP} tutorial is provided in~\cite{bip_tutorial}.
} It is important to clearly distinguish the \gls{BIP} features
from those of conventional programming languages, such as
C++ or Java.  First of all, being a framework, \gls{BIP} differs
from object-oriented programming, inasmuch as it relies on
the inversion of control principle: components notify the
BIP engine about possible transitions, then relinquish
control and wait for the engine to tell them which
transition to execute.  It is possible, but not necessary,
to associate calls to external functions to component
transitions.  The state of a \gls{BIP} component cannot be
directly modified by any other component.  The ports of BIP
components represent neither input/output data, nor methods
to be invoked.  Instead, a port denotes the \emph{event}
that occurs when the component executes a transition whereof
this port is the label.  Thus, a connector enforces
synchronisation of such events.  The \gls{BIP} engine decides
which components are to execute which transitions, based on
the full information about the connectors and priorities in
the system. Robustness of the approach is achieved via separation of concerns of software components and intrinsic capability of the framework to isolate error states. 

\subsubsection{Atoms}

BIP systems are assembled from atomic components (atoms),
corresponding to concurrent processes, such as control
algorithms, monitors, bus and memory drivers \etc Atoms have
disjoint state spaces.  An atom is defined by the
corresponding sets of ports, states, transitions, data
variables and update functions associated to transitions.
Simple functions\mdash as in the example below\mdash can be
specified directly, whereas more complex ones can be defined
as external C/C++ code.

\begin{figure}
	\centering 
	\resizebox{\columnwidth}{!}{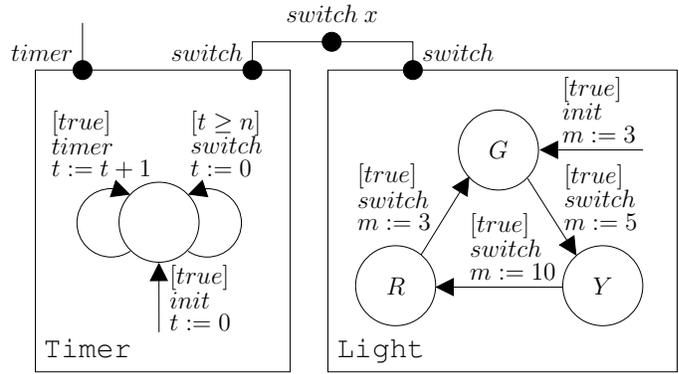}
	\caption{Traffic light in BIP}
	\label{fig:ex1}
\end{figure}

Figure~\ref{fig:ex1} shows a simple traffic light controller
system modeled in BIP.  It is composed of two atomic
components \texttt{Timer} and \texttt{Light}, modelling,
respectively, a timer and the light switching behaviour.
The \texttt{Timer} atom has one state with two self-loop
transitions.  The incoming arrow, labeled \emph{init},
denotes the initialisation event.  It is guarded by the
constant predicate $\true$ and has an associated update
function $t:=0$, which initialises the data variable $t$,
used to keep track of the time spent since the last change
of color.  The \texttt{Light} atom determines the color of
the traffic light and the duration that the light must stay
in one of the three states, corresponding to the three
colors.

\subsubsection{Connectors}

The system in \fig{ex1} has two connectors: a singleton
connector with one port $\mathit{timer}$ and no data
transfer and a binary connector, synchronising the ports
$switch$ of the two components.  The first, singleton
connector is necessary, since, in BIP, only ports that
belong to at least one connector can fire.  The second
connector synchronises the ports $\mathtt{Timer}.switch$ and
$\mathtt{Light}.switch$; it has an exported port, also
called $switch$, and an associated variable $x$ used for the
data transfer.

Connectors define sets of interactions based on the
synchronisation attributes of the connected ports, which may
be either \emph{trigger} or \emph{synchron}
(\fig[a]{connectors}).  If all connected ports are
synchrons, then synchronisation is by \emph{rendezvous}, \ie
the defined interaction may be executed only if all the
connected components allow the transitions of those ports
(\fig[b]{connectors}).  If a connector has at least one
trigger, the synchronisation is by \emph{broadcast}\footnote{
	Although we use the term ``broadcast'' by analogy with
	message passing\mdash trigger ports initialise
	interactions, whereas synchrons join if they are
	enabled\mdash, connectors synchronise ports\mdash no
	messages passing is involved.
}, \ie the
allowed interactions are all non-empty subsets of the
connected ports comprising at least one of the trigger ports
(\fig[b]{connectors}).  More complex connectors can be built
hierarchically (\fig[c]{connectors}).

\usetikzlibrary{shapes,arrows,decorations.markings}
\usetikzlibrary{backgrounds}
\usetikzlibrary{fit}
\usetikzlibrary{positioning}

\usetikzlibrary{calc}
\usetikzlibrary{automata,petri}
\usetikzlibrary{plotmarks}
\usetikzlibrary{decorations}
\usetikzlibrary{plotmarks}
\usetikzlibrary{patterns}
\usetikzlibrary{decorations.pathreplacing}

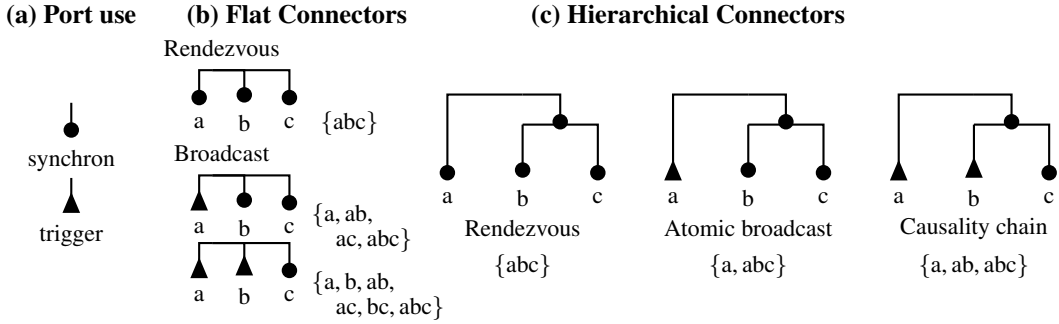
\begin{figure*}[t]
\noindent
\centering
\begin{tikzpicture}[shorten >=1pt,node distance=.7cm,>=stealth'
  ,initial text=
  ,every state/.style={draw=black,thick}
  ,group/.style = {draw=black,thick,rectangle, minimum width=2.5cm}
  ,port/.style = {font=\small}
  ,legend/.style = {font=\bf}
]

\node[port] (start) {};
\node[port, node distance=5.5cm] (h) [left of=start]{};

\node[legend](fltlgd) at ($(h.north)+(1,1.3cm)$) {(b) Flat Connectors};

\node[port, node distance=.3cm] (ar) [left of=h]{a};
\node[port, node distance=.3cm] (br) [right of=h]{b};
\node[port, node distance=.6cm] (cr) [right of=br]{c};
\draw [style=-*, thick]($(h.west)+(0,.7cm)$)-|(ar.north);
\draw [style=-*, thick] ($(h.west)+(0,.7cm)$) -|(br.north);
\draw [style=-*, thick] ($(h.west)+(0,.7cm)$) -|(cr.north);

\node[port, node distance=1.4cm] (h2) [below of=h]{};
\node[port, node distance=.3cm] (abr) [left of=h2]{a};
\node[port, node distance=.3cm] (bbr) [right of=h2]{b};
\node[port, node distance=.6cm] (cbr) [right of=bbr]{c};
\draw [style=-triangle 45 reversed, thick]($(h2.west)+(0,.7cm)$)-|(abr.north);
\draw [style=-*, thick] ($(h2.west)+(0,.7cm)$) -|(bbr.north);
\draw [style=-*, thick] ($(h2.west)+(0,.7cm)$) -|(cbr.north);

\node[port, node distance=.9cm] (h3) [below of=h2]{};
\node[port, node distance=.3cm] (adbr) [left of=h3]{a};
\node[port, node distance=.3cm] (bdbr) [right of=h3]{b};
\node[port, node distance=.6cm] (cdbr) [right of=bdbr]{c};
\draw [style=-triangle 45 reversed, thick]($(h3.west)+(0,.7cm)$)-|(adbr.north);
\draw [style=-triangle 45 reversed, thick] ($(h3.west)+(0,.7cm)$) -|(bdbr.north);
\draw [style=-*, thick] ($(h3.west)+(0,.7cm)$) -|(cdbr.north);

\node[port, node distance=1.7cm] (rndv) [right of=h]{\{abc\}};
\node[port, node distance=2.2cm] (brdc) [right of=h2]{\begin{tabular}{p{2cm}}\{a,\,ab,\,\\\quad ac,\,abc\}\end{tabular}};
\node[port, node distance=2.2cm] (dbrdc) [right of=h3]{\begin{tabular}{p{2cm}}\{a,\,b,\,ab,\,\\\quad ac,\,bc,\,abc\}\end{tabular}};

\node[port, node distance=1cm] (rndvs) [above of=h]{Rendezvous};
\node[port, node distance=1cm] (brdcs) [above of=h2]{Broadcast};

\node[port, fit=(rndvs)(dbrdc)](flat){};

\node[port, node distance=2cm](hll) [left of=h] {};
\node[legend](uselgd) at ($(hll.north)+(0,1.3cm)$) {(a) Port use};
\node[port, node distance=.5cm] (rndv2) [below of=hll]{synchron};
\node[port, node distance=1.5cm] (brdc2) [below of=hll]{trigger};
\draw [style=-*, thick] ($(rndv2.north)+(0,.5cm)$) -| (rndv2.north);
\draw [style=-triangle 45 reversed, thick] ($(brdc2.north)+(0,.5cm)$) -| (brdc2.north);

\node[port, fit=(rndv2)(brdc2)] (use) {};


\node[port, node distance=1.5cm](h4) [left of=start]{};

\node[legend](fltlgd) at ($(h4.north)+(2.2,1.3cm)$) {(c) Hierarchical Connectors};

\node[port, node distance=1cm] (ab) [below of=h4] {b};
\node[port, node distance=1cm] (ac) [right of=ab]{c};
\node[port, node distance=1cm] (aa) [left of=ab]{a};

\node[port, node distance=1.4cm] (leg4) [below of=h4] {Rendezvous};
\node[port, node distance=.5cm] (leg5) [below of=leg4] {\{abc\}};

\draw [style=-*, thick] ($(h4.south)+(0,.1cm)$) -- ++(right:.5cm) -| (ac.north);
\draw [style=-*, thick]  ($(h4.south)+(0,.1cm)$) -| (ab.north);
\draw [style=-*, thick] ($(h4.south)+(0,.5cm)$) -- ++(right:.2cm) -| ($(h4.south)+(+.5,0cm)$);
\draw [style=-*, thick]  ($(h4.south)+(0,.5cm)$) -| (aa.north);

\node[port, fit=(h4)(ab)(ac)(aa)(leg4)(leg5)](rdv){};


\node[port, node distance=1.5cm](baa) [right of=start]{};
\node[port, node distance=1cm] (bb) [below of=baa]{b};
\node[port, node distance=1cm] (bc) [right of=bb]{c};
\node[port, node distance=1cm] (ba) [left of=bb]{a};

\draw [style=-*, thick] ($(baa.south)+(0,.1cm)$) -- ++(right:.5cm) -| (bc.north);
\draw [style=-*, thick]  ($(baa.south)+(0,.1cm)$) -| (bb.north);
\draw [style=-*, thick] ($(baa.south)+(0,.5cm)$) -- ++(right:.2cm) -| ($(baa.south)+(+.5,0cm)$);
\draw [style=-triangle 45 reversed, thick]  ($(baa.south)+(0,.5cm)$) -| (ba.north);

\node[port, node distance=1.4cm] (leg6) [below of=baa] {Atomic broadcast};
\node[port, node distance=.5cm] (leg7) [below of=leg6] {\{a,\,abc\}};

\node[port, fit=(baa)(bb)(bc)(ba)(leg6)(leg7)](abrd){};


\node[port, node distance=4.5cm](caa) [right of=start]{};
\node[port, node distance=1cm] (cb) [below of=caa]{b};
\node[port, node distance=1cm] (cc) [right of=cb]{c};
\node[port, node distance=1cm] (ca) [left of=cb]{a};

\draw [style=-*, thick] ($(caa.south)+(0,.1cm)$) -- ++(right:.5cm) -| (cc.north);
\draw [style=-triangle 45 reversed, thick]  ($(caa.south)+(0,.1cm)$) -| (cb.north);
\draw [style=-*, thick] ($(caa.south)+(0,.5cm)$) -- ++(right:.2cm) -| ($(caa.south)+(+.5,0cm)$);
\draw [style=-triangle 45 reversed, thick]  ($(caa.south)+(0,.5cm)$) -| (ca.north);

\node[port, node distance=1.4cm] (leg8) [below of=caa] {Causality chain};
\node[port, node distance=.5cm] (leg9) [below of=leg8] {\{a,\,ab,\,abc\}};

\node[port, fit=(caa)(cb)(cc)(ca)(leg8)(leg9)](ccha){};








\end{tikzpicture}

\caption{Flat and hierarchical BIP connectors}
\label{fig:connectors}
\end{figure*}

In general, a connector description consists of three parts:
\begin{enumerate}
	\item A control part specifying a set of ports to be
	synchronised\mdash at most one per atomic component\mdash
	and, optionally, a single exported port. The latter can be
	used as a usual port in higher level connectors.
	
	\item A dataflow part specifying the computation associated
	with the interaction. The computation can affect variables
	associated with the ports. It consists of an upstream
	computation followed by a downstream computation.
	
	\item A Boolean guard determining the enabledness of an
	interaction depending on the values of the provided data:
	the interaction is only enabled if the data provided by
	the components satisfies the
	guard.
\end{enumerate}

The guard of the $\mathtt{Timer}.switch -
\mathtt{Light}.switch$ connector in \fig{ex1} is the
constant predicate $true$, the upward and downward dataflows
are defined, respectively, by the assignments $x :=
\mathtt{Timer}.m$ and $\mathtt{Light}.n := x$.  Thus, upon
each synchronisation, \texttt{Light} informs \texttt{Timer}
about the amount of time to spend in the next state, by
transferring the value of $\mathtt{Timer}.m$ into
$\mathtt{Light}.n$.  In hierarchical connectors, the
separation of the dataflow into upward- and downward parts
allows the data provided by the upward flow to be modified
in the higher levels of the connector hierarchy before being
transferred downwards.

\subsubsection{Priority}

Notice that, when $t \geq n$, both transitions, $timer$ and
$switch$, of the \texttt{Timer} atom are enabled.  Since all
other guards in the system are constant predicates $true$,
this means that both connectors can fire.  Imposing the
priority $timer < switch$ resolves this choice, so that
switching is performed whenever possible.  In general, it is
not necessary to impose priorities in all conflict
situations: according to the \gls{BIP} semantics, one of the
enabled maximal priority interactions is chosen
non-deterministically~\cite{BliSif08-acp-tc}.

\subsubsection{Compounds}

Finally, compound components are composed of sets of
sub-components (atoms and/or compounds), connectors and
priorities.  A compound can export ports defined in
connectors in order to interact with other components in a
larger compound.

\subsection*{Property Enforcement\mdash Architectures}
\label{secn:architectures:bip}

A posteriori verification, \eg model checking, is
well-known to be limited by the combinatorial state-space
explosion problem.  In particular, among the verification
tools discussed in the previous section, only DFinder is
known to scale well to very large models.  This is due to
the fact that this tool performs compositional
over-approximation of the set of reachable states of the
model, instead of computing it precisely.

By-construction property enforcement is an alternative
approach that allows designer to circumvent this
limitation.  It consists in applying design patterns\mdash
that we call \emph{architectures}\mdash to restrict the
behaviour of a set of components so that the composed
behaviour meets a given property.  Depending on the
expressiveness of the glue operators, it may be necessary
to use additional coordinating components to satisfy the
property.

Architectures depict design principles, paradigms that can
be understood by all, allow thinking on a higher level and
avoiding low-level mistakes.  They are a means for ensuring
global properties characterising the coordination between
components\mdash correctness for free.  Using architectures
is key to ensuring trustworthiness and optimisation in
networks, OS, middleware, HW devices \etc

System developers extensively use libraries of reference
architectures ensuring both functional and non-functional
properties, for example fault-tolerant architectures,
architectures for resource management and Quality of Service control,
time-triggered architectures, security architectures and
adaptive architectures.  The proposed definition is general
and can be applied not only to hardware or software
architectures but also to protocols, distributed
algorithms, schedulers, \etc

Given a semantic domain\mdash \ie a class of component
behaviors\mdash $\cC$, an architecture is a partial
operator $A: \cC^n \rightarrow \cC$, imposing a
characteristic property $\Phi$.  It is defined by a glue
operator (a combination of interactions and priorities)
$gl$ and a finite set of coordinating components $\cD
\subset \cC$, such that:

\begin{itemize}
	\item $A$ transforms a set of components $C_1, \dots, C_n$
	into a composite component $A[C_1, \dots, C_n]= gl(C_1,
	\dots, C_n, \cD)$;
	
	\item $A[C_1, \dots, C_n]$ meets the characteristic
	property $\Phi$.
\end{itemize}

An architecture is a solution to a coordination problem
specified by $\Phi$, using a particular set of interactions
specified by $gl$.  It is a partial operator, since the
interactions of $gl$ should match actions of the composed
components.
\begin{figure}
	\includegraphics[width=\columnwidth]{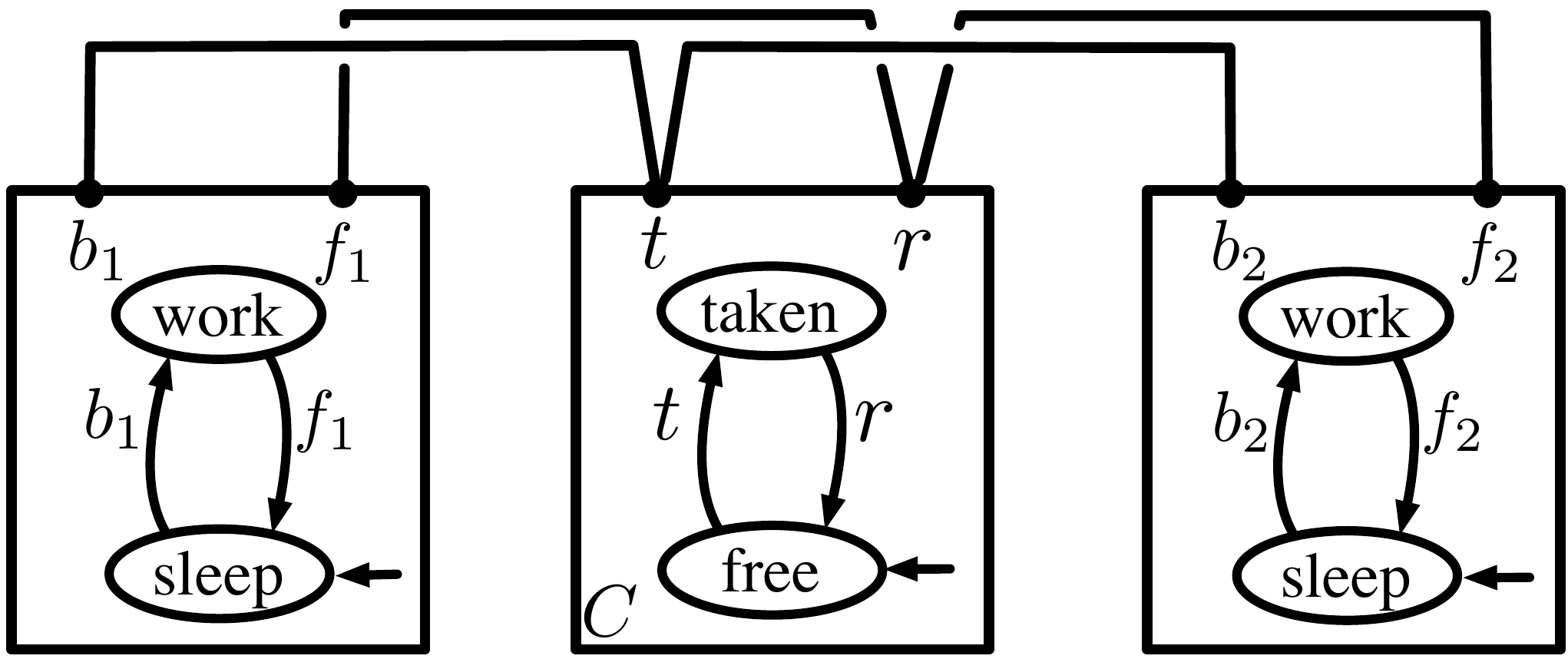}
	\caption{Mutual exclusion model in BIP}
	\label{fig:ex:mutex}
\end{figure}

Figure~\ref{fig:ex:mutex} shows a simple \gls{BIP} model for
mutual exclusion between two tasks.  It has two components
modeling the tasks and one coordinator component $C$.  The
four binary connectors synchronise each of the actions
$b_1$, $b_2$ (resp. $f_1$, $f_2$) of the tasks with the
action $t$, for ``take'', (resp. $r$, for ``release'') of
the coordinator.

\begin{figure}
	\includegraphics[width=\columnwidth]{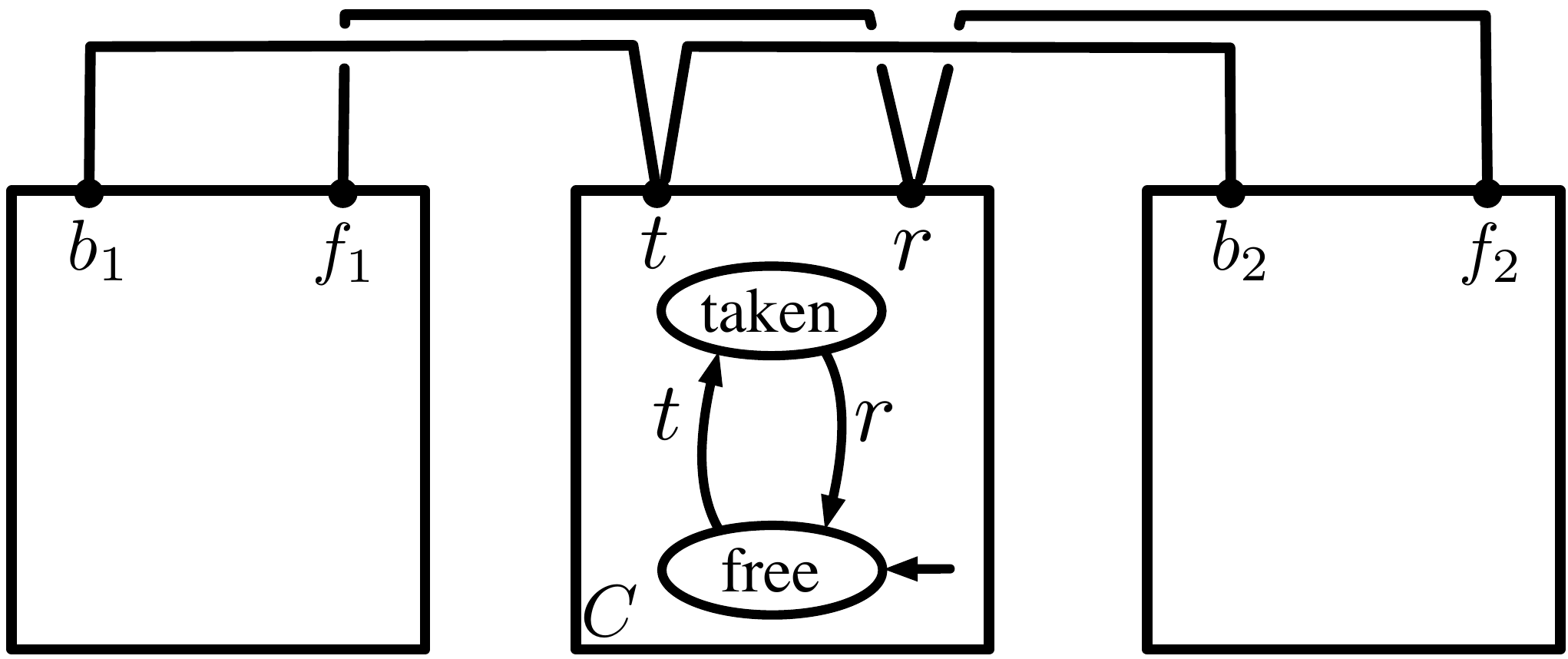}
	\caption{Mutual exclusion architecture}
	\label{fig:arch:mutex}
\end{figure}

An architecture can be viewed as a \gls{BIP} model, where some of
the atomic components are considered as \emph{coordinators},
while the rest are \emph{parameters}.  When an architecture
is applied to a set of components, these components are used
as \emph{operands} to replace the parameters of the
architecture.  Clearly, operand components must refine the
corresponding parameter ones\mdash in that sense, parameter
components can be considered as \emph{types}.\footnote{%
	The precise definition of the refinement relation is
	beyond the scope of this paper.  
} Thus, \fig{arch:mutex} shows an architecture that enforces
the mutual exclusion property on any two components with
interfaces $\{b_1,f_1\}$ and $\{b_2,f_2\}$, satisfying the
following property: 
\begin{center}
	\parbox{0.9\columnwidth}{\em Once a component has executed
		the $f$ transition, it will not enter the critical
		section as long as it does not execute the $b$
		transition.}
\end{center}
For the
operand components in \fig{ex:mutex}, the critical
section is the state \texttt{work}.

Composition of architectures is based on an associative,
commutative and idempotent architecture composition operator
`$\oplus$'~\cite{AttieBBJS16-architectures-faoc}.  If two
architectures \cA${}_1$ and \cA${}_2$ enforce respectively
safety properties $\Phi_1$ and $\Phi_2$, the composed
architecture $\cA_1 \oplus \cA_2$ enforces the property
$\Phi_1 \land \Phi_2$, that is both properties are preserved
by architecture composition.  

Since architectures restrict the behavior of components they
are applied to, preservation of deadlock-freedom cannot, in
general, be guaranteed by construction.  Instead,
deadlock-freedom has to be verified a posteriori using
dedicated tools, such as DFinder.  The role of model
checking is reduced to demonstrating the correctness of
architectures \wrt their characteristic properties.  This
can usually be achieved by using relatively small models
combined with inductive reasoning techniques to formally
prove that, if a given architecture correctly enforces its
characteristic property on a certain number of operand
components, it will also correctly impose this property on
any number of components.

\label{secn:sw}

\section{Approach}
\label{secn:approach}

\label{secn:goals}

Our goal is to provide a framework for rigorous design and
implementation of control software for space missions.  Such framework
must have the following key properties.

\subsubsection{Concurrency} 
Control software and, in particular, that required for space missions
is inherently concurrent, since space systems comprise large numbers
of elements, such as sensors, scientific instruments and other
sub-systems that operate in real time,
sharing resources, such as communication buses and memories.

\subsubsection{Modularity}
Software driving the operation of these elements must be designed in a
modular fashion, with the structure of the software closely mimicking
that of the system.  On one hand, this allows separate validation of
each software component and, on the other hand, greatly improves
reconfigurability of the system.

\subsubsection{Separation of concerns}
In order to ensure robustness of the software system, designers must
have the capability to identify all error states and ensure that
corresponding corrective actions are properly defined.  To this end,
the above modularity requirement is strengthened by further requiring
that the state information be clearly exhibited in the software model.

\subsubsection{Expressiveness}
The design framework should be sufficiently expressive to allow the
development of complete software systems without the need for complex
manual integration.  
The state space of
such systems grows exponentially with the number of components.

\subsubsection{Visual editing}
In order to further reduce cognitive complexity and allow designers to
have a global understanding of the software system on the levels of
abstraction appropriate for each design task, the design framework must
provide a visual representation and editing front-end allowing
designer to browse the model of the software in an intuitive way and
structuring the model information that is presented at any given time.

\subsubsection{Correctness guarantees}
It is fundamental for the design of mission-critical software to have
the possibility of establishing the correctness of the software model
with respect to the mission requirements and, in particular, the
satisfaction of safety properties and deadlock-freedom.  To this end,
the design framework must provide either verification tools capable of
analysing the designed models, or synthesis tools automatically
deriving component behaviour from high-level specifications, or\mdash
more pragmatically\mdash a combination of both.

\subsubsection{Formal operational semantics}  
Guarantees of correctness can only be provided if the modelling
formalism underlying the design framework has a formally defined
operational semantics.  In the absence of such operational semantics,
formal reasoning about the system is not possible and the only options
for software validation available at design time are testing and
simulation.  
At CERN the use of formal
verification techniques has allowed finding bugs even in relatively
simple, well-tested components~\cite{Borja}.

\subsubsection{Automatic code generation}
Lastly, to ensure that the properties
demonstrated to hold on the model of the software system remain valid
in the final implementation, it is highly desirable that the
executable code be automatically generated from the model by a
semantics-preserving transformation (to ensure forward and backward translation).


\section{Implementation}
\label{secn:implementation}

\begin{figure*}[t]
  \centering 
  \includegraphics[scale=0.4]{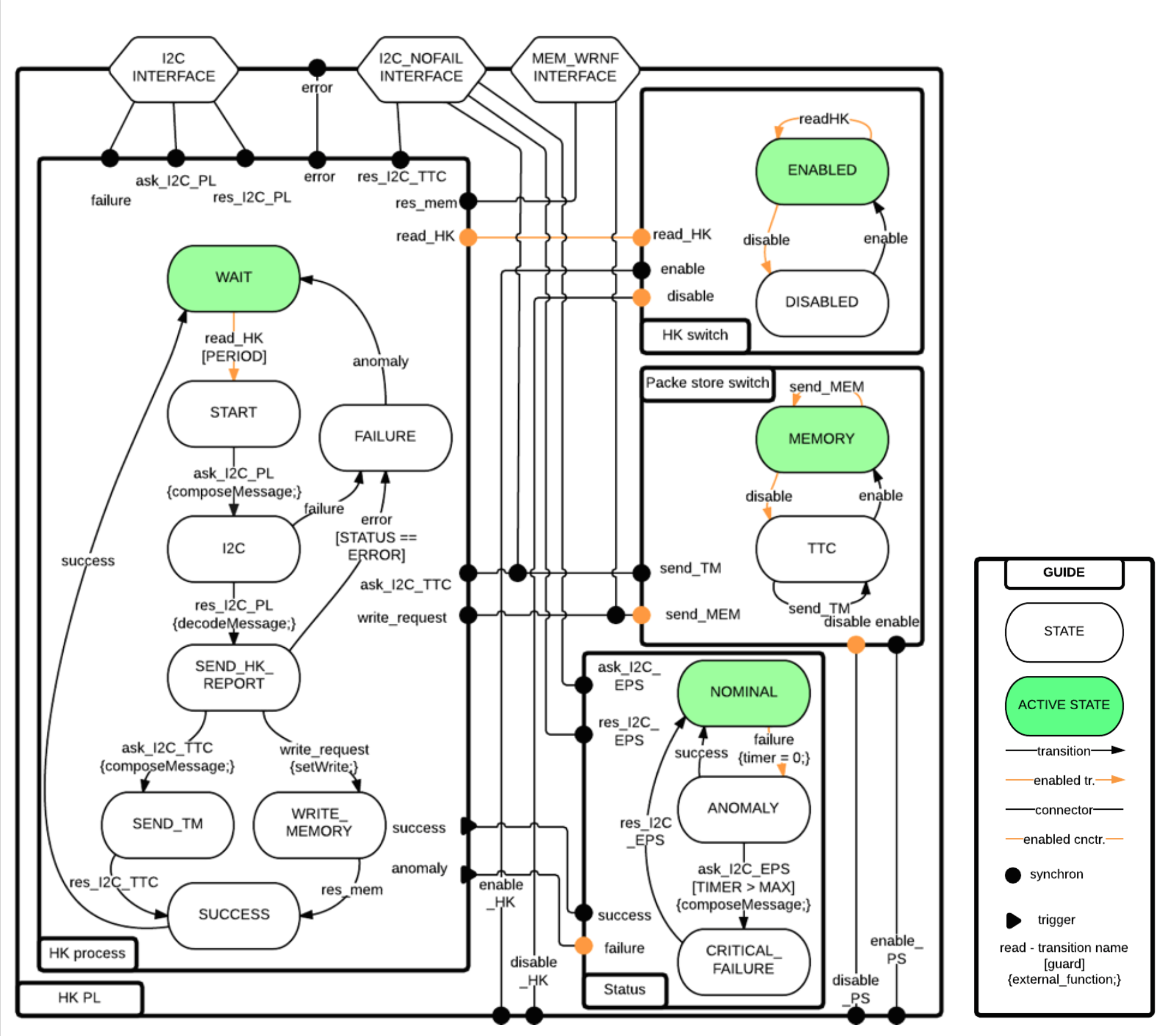}	
  \caption{The BIP model of the payload housekeeping readout block}
  \label{payload_hk}

  \parbox{\textwidth}{The ports comprising external
      interfaces of the compound component are located on
      the boundaries of the block.  Transitions are executed
      on external signals (\eg \texttt{read\_HK},
      \texttt{enable}, \texttt{disable}, \texttt{send\_TM}
      \etc Active states are highlighted in green, whereas
      enabled ports and connectors are highlighted in
      orange. The block is shown in its \texttt{WAIT} mode.}
\end{figure*}

The full design workflow (\ie requirements analysis, logic design,
BIP code development, compilation to microprocessor and testing) was
completed~\cite{MarcoMasterThesis}, thereby demonstrating the feasibility of the \gls{BIP} concept 
application to nanosatellties. An example of the Payload Housekeeping element
is  shown in Figure~\ref{payload_hk}.  

The full model had 56 atoms and did not fit into the available memory on the satellite Engineering Model.  Therefore, the full tool chain was implemented for a reduced software model, containing 19 atoms and 60 connectors.  It should be noted that the reduction of the model was greatly simplified by the strong modularity provided by the BIP framework. At the moment, the feasibility study has implemented a small part of the overall system. Full system will be implemented in the future on a system with more available memory. 

\begin{figure}[t]
  \centering 
  \includegraphics[width=8cm]{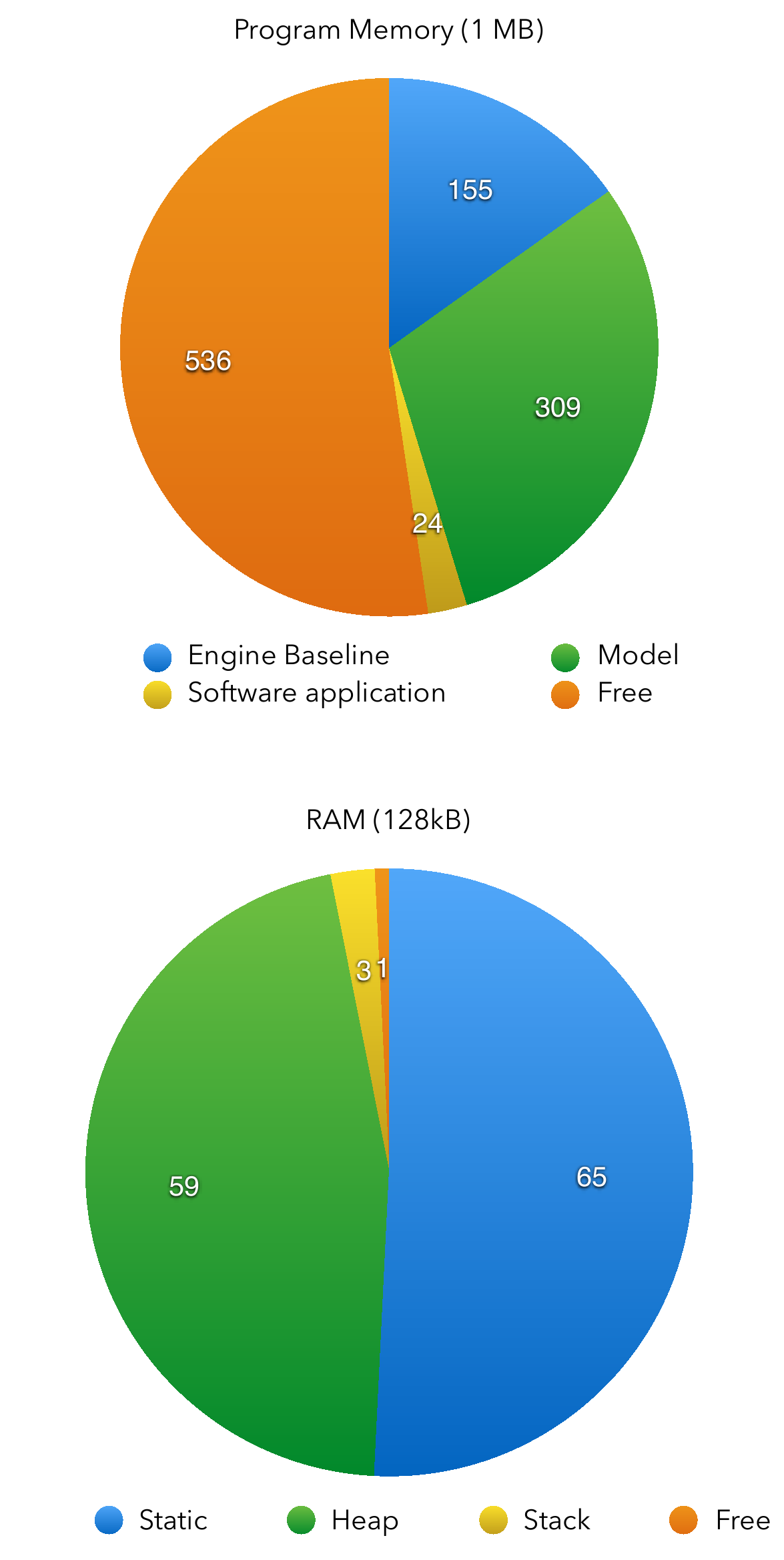}	
  \caption{Memory footprint of the reduced \glsentrytext{BIP} model}
  \label{MarcoRAMfootprint}
  
  \settowidth{\templength}{{\bf Figure 6. } Memory footprint of the reduced \glsentrytext{BIP} model}

  \parbox{\templength}{The model was compiled for the Energy Micro EFM32 processor (a Cortex-M3 running at 48~MHz). The top chart shows the partition of the total allocated memory: note that there is still 536~kB available. The bottom chart is for allocation in the RAM, which is the source of the main limitation on the size of the \gls{BIP} model.}
\end{figure}

The model represented the interaction of the CDMS (Control and Data Management Subsystem) with the Payload and the Attitude Determination and Control (ADCS) subsystems.  Model time step on the microprocessor was 6~ms, which satisfied the requirement for 0.5~Hz processing step on the satellite bus.  The main challenges were 1)~the need to perform static allocation of memory, 2)~the large RAM footprint of the generated code and 3)~the necessity of compiling the BIP engine and the generated code with a custom ARM compiler. 
The memory footprint of the generated code is shown in Figure~\ref{MarcoRAMfootprint}. The amount of RAM (128MB) on the Cortex-A3 processor is the main constraint, limiting the number of atoms and connectors (and hence the complexity of the model) that can be used simultaneously. Only static allocation is available, therefore it is impossible to reallocate data to main memory. This restriction is easily lifted on more powerful platforms. Selected architecture was used due to its low power consumption and, more generally, the restrictive power regime on a 1U CubeSat. Complex projects with university CubeSats are now moving fast towards 3U and 6U platforms. We expect that future platforms will be more powerful and therefore can accommodate full model implementation. 

 Several strategies were considered in~\cite{Sikiaridis2016} for the validation of the model,
 with mixed success. Application of verification
 tools proved too complex on the complete satellite model with full detail for the following two reasons. First,
 the available prototype verification tools only support a reduced \gls{BIP} syntax.  In particular, compound components are not supported and the external C/C++ code, which is not analysed as part of the verification process, has to be removed manually.  The second, intrinsic reason of the verification complexity is the large size of the state-space that must be analysed, which calls for non-trivial abstractions and for compositional approaches that are not available in the current state of the art, for general verification. (As discussed below, verification of specific properties, such as deadlock freedom, can be performed compositionally.)  Further analysis in~\cite{MarcoMasterThesis,Sikiaridis2016} has suggested that \emph{by-construction} validity could be achieved by applying specific model development rules for the design of the BIP model of the software. 

We have analysed the model developed
by~\cite{MarcoMasterThesis} to identify recurring design
patterns.  These patterns were formalised as BIP
architectures (see \secn{architectures:bip}) and used in
\cite{facs16cubeth} to design a new BIP model for CubETH
on-board software from scratch.  More specifically,
starting with a small set of atoms, which realise the simple
key functionality of the software, we have systematically
applied the above architectures to enforce safety properties
associated to the requirements formulated for the on-board
software.  Since the safety properties imposed by
architectures are preserved by architecture
composition~\cite{AttieBBJS16-architectures-faoc}, all
properties that we have associated to the CubETH
requirements are satisfied by construction by this
latter model.

Architectures enforce properties by restricting the joint
behaviour of the operand components (\secn{architectures:bip}).  Therefore, combined
application of architectures can generate deadlocks.  We
have used the \texttt{DFinder} tool~\cite{dfinder} to
verify deadlock-freedom of the case study model.
\texttt{DFinder} applies compositional verification on BIP
models by over-approximating the set of reachable states 
and checking symbolically that the intersection of the
obtained set and the set of deadlock states is empty.
This approach allows DFinder to analyse very large models.  The tool is
sound, but incomplete: due to the above mentioned
over-approximation it can produce false positives, \ie
potential deadlock states that are unreachable in the
concrete system.  However, our case study model was shown to
be deadlock-free without any potential deadlocks.  Thus, no
additional reachability analysis was needed.

\section{Discussion} 

The main drawback of the work in~\cite{MarcoMasterThesis} resides in the fact that
the model was initially drawn by hand using LucidCharts\footnote{%
	\url{https://www.lucidchart.com/}
}, 
rendering impossible the translation of diagrams into any other
language.  Hence, \gls{BIP} code reproducing the logic in the
diagrams had to be written manually, with the inherent risk of
introducing errors and reduced
maintainability of the code. 
However, the nature of \gls{BIP}
is graphical, due to the component-based philosophy adopted by the framework, the use of Labelled Transition Systems and connectors to define, respectively, component behaviour and synchronisations.  Therefore, we have made the following two attempts to address this problem.

The first attempt~\cite{Pittet2015} consisted in re-encoding this model
in SysML, using activity diagrams. This attempt failed due to the absence of versatile support for 
synchronization modeling in SysML.
We have subsequently attempted to develop a visual editor based on the Sirius\footnote{\url{http://www.eclipse.org/sirius/}} plugin for
Eclipse, directly using the \gls{BIP} ECORE meta-model
\cite{Ilievski2016}. 

Eclipse Sirius is an open-source software project of the Eclipse
Foundation. This technology allows creating custom graphical modeling
workbenches by leveraging the Eclipse Modeling technologies. Sirius is
mainly used to design complex systems (industrial systems or IT
applications). The first use case was 
Capella\footnote{\url{http://www.polarsys.org/capella/}}, a Systems
Engineering workbench contributed to the Eclipse Working Group
PolarSys in 2014 by Thales. 

A prototype modeler for specifying \gls{BIP} systems using the
Eclipse Sirius tool was developed, providing a set of diagrams with all constituent elements of \gls{BIP} models, such as compounds, atoms, ports and connectors.  The modeler generates an XMI file in the format accepted by the \gls{BIP}
code generator, thus allowing graphical design of a working \gls{BIP} system.

However, the visual interface of the modeler was unsatisfactory, significantly departing from the graphical conventions commonly adopted in \gls{BIP}.  This is due to the
structural incompatibility of the current Ecore meta-model used in the BIP framework with user-friendly visual
rendering.  
Thus, we conclude that a new visual editor design is necessary, which
would be based on a dedicated meta-model with a bi-directional
transformation into the \gls{BIP} meta-model.  

\section{Summary} 
In this work, we have demonstrated the feasibility of the \gls{BIP}
approach for the development of flight software.  Restrictions of the
Cortex-M3 processor have forced us to reduce the model in order to fit
it into the memory available on the CubeSat.  However, the
demonstration of the reduced model on the CubeSat board was a success.

We found it extremely useful to have a capability to verify the design
logic before compiling to hardware.  Although we did not succeed in
formally verifying the complete model, due to a combination of
intrinsic (model complexity) and extrinsic (tool limitations) factors,
the structure of the satellite software makes this model readily
amenable for decomposition into a number of parts, each of which can
be verified individually.  Furthermore, the application of the \gls{BIP}
architecture-based design approach has allowed us to design a similar
model, where all the safety properties are enforced by construction.
This approach only requires the model to be verified for deadlock
freedom, which we managed to achieve using the DFinder tool from the
\gls{BIP} toolset.  (It should be noted, however, that any properties
that are not enforced by construction would still have to be verified
separately.)

\section{Future work} 
This work has shown that, for relatively small missions, such as
CubSats, \gls{BIP} can be used to design \emph{complete} on-board
software.  In the context of larger missions, where design of software
components is often delegated to sub-contractors, a key difficulty
consists in the integration of such components.  The
\gls{osra}~\cite{SAVOIR-FAIRE} standardization initiative aims at
addressing this difficulty by defining a common component model.  In
this context, the most straightforward application of the \gls{BIP}
framework is the development and automatic code generation for
\gls{osra} components.  To achieve this goal, the \gls{BIP} code
generation must be adapted in such a manner as to provide the
interfaces defined by the \gls{osra}.  Furthermore, availability of
\gls{BIP} models\mdash developed in the framework of the ESA
``Catalogue of Software and System Properties'' project\mdash for the
\gls{osra} constituent elements, allows the use of the \gls{BIP}
framework for co-simulation and co-validation of several
\emph{heterogeneous} components, whereof parts can be designed in
\gls{BIP}, while others\mdash using alternative languages and
modelling frameworks, such as C++ or Mathlab/Simulink.

We have identified a number of developments necessary to facilitate OBSW design using the \gls{BIP} approach. A project is currently under way to 
prepare a graphical user interface, necessary to avoid manual transfer of diagrams to code. It is also necessary to study the power consumption overhead when using the precompiled model. 

Finally, an important part of the future work is to design a set of design patterns and validation rules to improve quality of the software architecture. 

\acknowledgments The authors would like to thank the Master and PhD
students who have contributed to this project.  We are also grateful
to the \'Ecole polytechnque f\'ed\'erale de Lausanne for supporting it.

\bibliographystyle{IEEEtran}
\bibliography{bip_refs,biblio,biplan_refs}

\begin{thebibliography}{10}
\providecommand{\url}[1]{#1}
\csname url@samestyle\endcsname
\providecommand{\newblock}{\relax}
\providecommand{\bibinfo}[2]{#2}
\providecommand{\BIBentrySTDinterwordspacing}{\spaceskip=0pt\relax}
\providecommand{\BIBentryALTinterwordstretchfactor}{4}
\providecommand{\BIBentryALTinterwordspacing}{\spaceskip=\fontdimen2\font plus
\BIBentryALTinterwordstretchfactor\fontdimen3\font minus
  \fontdimen4\font\relax}
\providecommand{\BIBforeignlanguage}[2]{{%
\expandafter\ifx\csname l@#1\endcsname\relax
\typeout{** WARNING: IEEEtran.bst: No hyphenation pattern has been}%
\typeout{** loaded for the language `#1'. Using the pattern for}%
\typeout{** the default language instead.}%
\else
\language=\csname l@#1\endcsname
\fi
#2}}
\providecommand{\BIBdecl}{\relax}
\BIBdecl

\bibitem{Brandon2013}
C.~Brandon and P.~Chapin, \emph{{Reliable Software Technologies – Ada-Europe
  2013}}, ser. Lecture Notes in Computer Science, H.~B. Keller,
  E.~Pl{\"{o}}dereder, P.~Dencker, and H.~Klenk, Eds.\hskip 1em plus 0.5em
  minus 0.4em\relax Berlin, Heidelberg: Springer Berlin Heidelberg, 2013, vol.
  7896.

\bibitem{sysmlpaper}
S.~C. Spangelo, J.~Cutler, L.~Anderson, E.~Fosse, L.~Cheng, R.~Yntema,
  M.~Bajaj, C.~Delp, B.~Cole, G.~Soremekum, and D.~Kaslow, ``{Model based
  systems engineering (MBSE) applied to Radio Aurora Explorer (RAX) CubeSat
  mission operational scenarios},'' in \emph{2013 IEEE Aerospace
  Conference}.\hskip 1em plus 0.5em minus 0.4em\relax IEEE, mar 2013, pp.
  1--18.

\bibitem{Dathathri2016}
S.~Dathathri, S.~C. Livingston, R.~M. Murray, and L.~J. Reder, ``{Interfacing
  TuLiP with the JPL Statechart Autocoder : Initial progress toward synthesis
  of flight software from formal specifications},'' in \emph{IEEE AeroSpace},
  2016.

\bibitem{esst4bip}
S.~Bliudze, A.~Cimatti, M.~Jaber, S.~Mover, M.~Roveri, W.~Saab, and W.~Qiang,
  ``Formal verification of infinite-state {BIP} models,'' in \emph{Proceedings
  of the 13th {I}nternational {S}ymposium on {A}utomated {T}echnology for
  {V}erification and {A}nalysis}, ser. Lecture Notes in Computer Science,
  B.~Finkbeiner, G.~Pu, and L.~Zhang, Eds., vol. 9364.\hskip 1em plus 0.5em
  minus 0.4em\relax Springer, 2015, pp. 326--343.

\bibitem{Ivanov2015}
A.~B. Ivanov, L.~Masson, S.~Rossi, F.~Belloni, N.~Mullin, R.~Wiesendanger,
  M.~Rothacher, C.~Hollenstein, B.~Mannel, D.~Willi, M.~Fisler, P.~Fleischman,
  H.~Mathis, M.~Klaper, M.~Joss, and E.~Styger, ``{CubETH}: Nano-satellite
  mission for orbit and attitude determination using low-cost {GNSS}
  receivers,'' in \emph{66th International Astronautical Congress}.\hskip 1em
  plus 0.5em minus 0.4em\relax Jerusalem, Israel: International Astronautical
  Federation, IAF, 2015.

\bibitem{Rossi20151513}
\BIBentryALTinterwordspacing
S.~Rossi, A.~Ivanov, G.~Soudan, V.~Gass, C.~Hollenstein, and M.~Rothacher,
  ``{CubETH} magnetotorquers: Design and tests for a cubesat mission,''
  \emph{Advances in the Astronautical Sciences}, vol. 153, pp. 1513--1530,
  2015. [Online]. Available:
  \url{https://www.scopus.com/inward/record.uri?eid=2-s2.0-84968735754&partnerID=40&md5=bb518e947c8ab4ddbeac99ac1a755895}
\BIBentrySTDinterwordspacing

\bibitem{Rossi20151493}
\BIBentryALTinterwordspacing
S.~Rossi, A.~Ivanov, G.~Burri, V.~Gass, C.~Hollenstein, and M.~Rothacher,
  ``{CubETH} sensor characterization: Sensor analysis required for a cubesat
  mission,'' \emph{Advances in the Astronautical Sciences}, vol. 153, pp.
  1493--1512, 2015. [Online]. Available:
  \url{https://www.scopus.com/inward/record.uri?eid=2-s2.0-84968718629&partnerID=40&md5=aac75ea756db639e4bc94d5fbd24c473}
\BIBentrySTDinterwordspacing

\bibitem{ECSSEST40C}
ECSS, ``{ECSS-E-ST-40C: Software},'' ESA Requirements and Standards Division,
  Noordwijk, Netherlands, Tech. Rep., 2009.

\bibitem{Wilmot2016}
J.~Wilmot, L.~Fesq, and D.~Dvorak, ``{Quality Attributes for Mission Flight
  Software : A Reference for Architects},'' in \emph{IEEE AeroSpace}, Big Sky,
  MT, 2016, pp. 1--7.

\bibitem{main_bip}
A.~Basu, S.~Bensalem, M.~Bozga, J.~Combaz, M.~Jaber, T.-H. Nguyen, and
  J.~Sifakis, ``Rigorous component-based system design using the {BIP}
  framework,'' \emph{Software, IEEE}, vol.~28, no.~3, pp. 41--48, 2011.

\bibitem{BliSif08-acp-tc}
S.~Bliudze and J.~Sifakis, ``The algebra of connectors---structuring
  interaction in {BIP},'' \emph{{IEEE} Transactions on Computers}, vol.~57,
  no.~10, pp. 1315--1330, 2008.

\bibitem{MiSE14p25}
S.~Bliudze, A.~Mavridou, R.~Szymanek, and A.~Zolotukhina, ``Coordination of
  software components with {BIP}: Application to {OSGi},'' in \emph{Proceedings
  of the 6th International Workshop on Modeling in Software Engineering}, ser.
  MiSE 2014.\hskip 1em plus 0.5em minus 0.4em\relax New York, NY, USA: ACM,
  2014, pp. 25--30.

\bibitem{dfinder}
S.~Bensalem, A.~Griesmayer, A.~Legay, T.-H. Nguyen, J.~Sifakis, and R.~Yan,
  ``{D-Finder 2}: towards efficient correctness of incremental design,'' in
  \emph{Proceedings of the $3^{\mathrm{rd}}$ international conference on NASA
  Formal methods}, ser. NFM'11.\hskip 1em plus 0.5em minus 0.4em\relax Berlin,
  Heidelberg: Springer-Verlag, 2011, pp. 453--458.

\bibitem{bip_tutorial}
``{BIP} tutorial,''
  \url{http://www-verimag.imag.fr/TOOLS/DCS/bip/doc/latest/html/language.html}.

\bibitem{AttieBBJS16-architectures-faoc}
P.~Attie, E.~Baranov, S.~Bliudze, M.~Jaber, and J.~Sifakis, ``A general
  framework for architecture composability,'' \emph{Formal Aspects of
  Computing}, vol.~18, no.~2, pp. 207--231, Apr. 2016.

\bibitem{Borja}
B.~Fern\'andez~Adiego, D.~Darvas, E.~Blanco Vi\~nuela, J.-C. Tournier,
  S.~Bliudze, J.~O. Blech, and V.~M. Gonz\'alez~Su\'arez, ``Applying model
  checking to industrial-sized {PLC} programs,'' \emph{{IEEE} {T}ransactions on
  {I}ndustrial {I}nformatics}, vol.~11, no.~6, pp. 1400--1410, 2015.

\bibitem{MarcoMasterThesis}
M.~Pagnamenta, ``Rigorous software design for nano- and micro-satellites using
  {BIP} framework,'' Master's thesis, Space Center, EPFL, Sep. 2014.

\bibitem{Sikiaridis2016}
A.~Sikiaridis, ``{Definition and Implementation of Validation Strategies for a
  Nanosatellite Flight Control Software Model},'' eSpace, EPFL, Lausanne,
  Switzerland, Tech. Rep., 2016.

\bibitem{facs16cubeth}
A.~Mavridou, E.~Stachtiari, S.~Bliudze, A.~Ivanov, P.~Katsaros, and J.~Sifakis,
  ``Architecture-based design: A satellite on-board software case study,'' in
  \emph{Proceedings of the 13th International Conference on Formal Aspects of
  Component Software}, O.~Kuchnarenko and R.~Khosravi, Eds., 2016, to appear.

\bibitem{Pittet2015}
J.-N. Pittet, ``{CubETH: Onboard software design with SysML},'' eSpace, EPFL,
  Lausanne, Switzerland, Tech. Rep., 2015.

\bibitem{Ilievski2016}
V.~Ilievski, ``{Implemtation of visual interface for BIP programming in Eclipse
  Sirius environment},'' RISD, Lausanne, Tech. Rep., 2016.

\bibitem{SAVOIR-FAIRE}
A.~Jung, M.~Panunzio, and J.-L. Terraillon, ``{SAVOIR-FAIRE} --- {On-board}
  software reference architecture,'' SAVOIR Advisory Group, Tech. Rep.
  TEC-SWE/09-289/AJ, Jun. 2010.

\end{thebibliography}

\thebiography
\begin{biographywithpic}
{Dr Anton B. Ivanov}{Anton2012IEEE}
is a scientist with the EPFL Space Center (eSpace) in Lausanne Switzerland. He is the project manager for the CubETH CubeSat project, study leader for the CHEOPS satellite and is responsible for the Minor in Space Technologies. After receiving his PhD in Planetary Science from Caltech in 2000, Dr Ivanov joined the Jet Propulsion Laboratory to contribute to Mars Global Surveyor, Mars Odyssey, Mars Express and Mars Science Laboratory projects. In 2007, Dr Ivanov joined Swiss Space Center to lead development of the Concurrent Design Facility.
\end{biographywithpic} 

\begin{biographywithpic}
{Dr Simon Bliudze}{SB} holds an MSc in Mathematics from the
St.\ Petersburg State University (Russia, 1998), an MSc in
Computer Science from Universit\'e Paris 6 (France, 2001)
and a PhD in Computer Science from \'Ecole Polytechnique
(France, 2006). He has spent two years at Verimag (Grenoble,
France) as a post-doc with Joseph Sifakis working on formal
semantics for the \gls{BIP} component framework. After three years
as a research engineer at CEA Saclay, France, he has joined
the Rigorous System Design Laboratory (RiSD) at EPFL.  Since
2014, Dr Bliudze is working on the aplication of \gls{BIP} to the
design of On-Board Software in collaboration with the EPFL
Space Center and in the Catalogue of System and Software
Properties (CSSP) project funded by hte European Space
Agency.
\end{biographywithpic}

\end{document}